\documentclass[twocolumn,twoside,slac_two]{revtex4}
\usepackage{graphicx}
\usepackage{fancyhdr}
\begin{document}

\title{Detection of TeV Gamma-Rays from extended sources with Milagro}

%

\author{P.~M.~Saz Parkinson for the Milagro Collaboration\footnote{R.~Atkins,  
W.~Benbow, 
D.~Berley, 
E.~Blaufuss, 
D.~G.~Coyne, 
T.~DeYoung, 
B.~L.~Dingus, 
D.~E.~Dorfan, 
R.~W.~Ellsworth, 
L.~Fleysher,
R.~Fleysher,
M.~M.~Gonzalez,
J.~A.~Goodman,
T.~J.~Haines,
E.~Hays,
C.~M.~Hoffman,
L.~A.~Kelley,
C.~P.~Lansdell,
J.~T.~Linnemann,
J.~E.~McEnery,
A.~I.~Mincer,
M.~F.~Morales,
P.~Nemethy,
D.~Noyes,
J.~M.~Ryan,
F.~W.~Samuelson,
P.~M.~Saz Parkinson,
A.~Shoup,
G.~Sinnis,
A.~J.~Smith,
G.~W.~Sullivan,
D.~A.~Williams,
M.~E.~Wilson,
X.~W.~Xu
and 
G.~B.~Yodh}}

\affiliation{Santa Cruz Institute for Particle Physics, University of California, 1156 High Street, Santa Cruz, CA 95064}
\begin{abstract}
The Milagro gamma-ray observatory employs a water Cherenkov detector to 
observe extensive air showers produced by high-energy particles impacting 
in the Earth's atmosphere. 
A 4800 m$^{2}$ pond instrumented with 723 8" PMTs detects Cherenkov light 
produced by secondary air-shower particles. An array of 175 4000 liter  
water tanks surrounding the central pond detector was recently added, 
extending the physical area of the Milagro observatory to 40,000 m$^{2}$ and 
substantially increasing the sensitivity of the detector. 
Because of its wide field of view and high duty cycle, Milagro is
ideal for monitoring the northern sky almost continuously ($>$90\% duty cycle) 
in the 100 GeV to 100 TeV energy range. Here we discuss the first detection 
of TeV gamma-rays from the inner Galactic plane region. We also report the 
detection of an extended TeV source coincident with the EGRET source 
3EG J0520+2556, as well as the observation of extended TeV emission from 
the Cygnus region of the Galactic plane.

\end{abstract}

\maketitle

\thispagestyle{fancy}


\section{THE MILAGRO OBSERVATORY}
\begin{figure}[t]
\centering
\includegraphics[width=70mm]{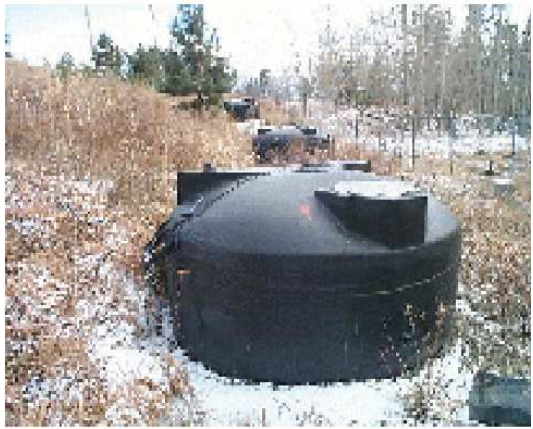}
\caption{One of 175 outriggers (recently added to Milagro).} \label{outrigger}
\end{figure}

Milagro is a TeV gamma-ray detector which uses the water Cherenkov technique to detect 
extensive air-showers produced by very high energy (VHE, $>$ 100 GeV) gamma rays as they 
interact with the Earth's atmosphere. Milagro is located in the Jemez Mountains of northern 
New Mexico, at an altitude of 2630 m,  has a field of view of $\sim$2 sr and a duty cycle 
greater than 90\%. The effective area of Milagro is a function of zenith angle and 
ranges from $\sim10 m^2$ at 100 GeV to $\sim10^5 m^2$ at 10 TeV. A sparse array of 175 
4000 liter water tanks, each containing an individual PMT, was recently added. These 
additional detectors, known as ``outriggers'' (Figure~\ref{outrigger}), extend the 
physical area of Milagro to 40,000 $m^2$, substantially increasing the sensitivity of the 
instrument and lowering the energy  threshold. The angular resolution is approximately 
0.75 degrees without the outriggers and 0.45 degrees with them.

\section{THE ALL-SKY SEARCH}
\begin{figure*}[t]
\centering
\includegraphics[width=170mm]{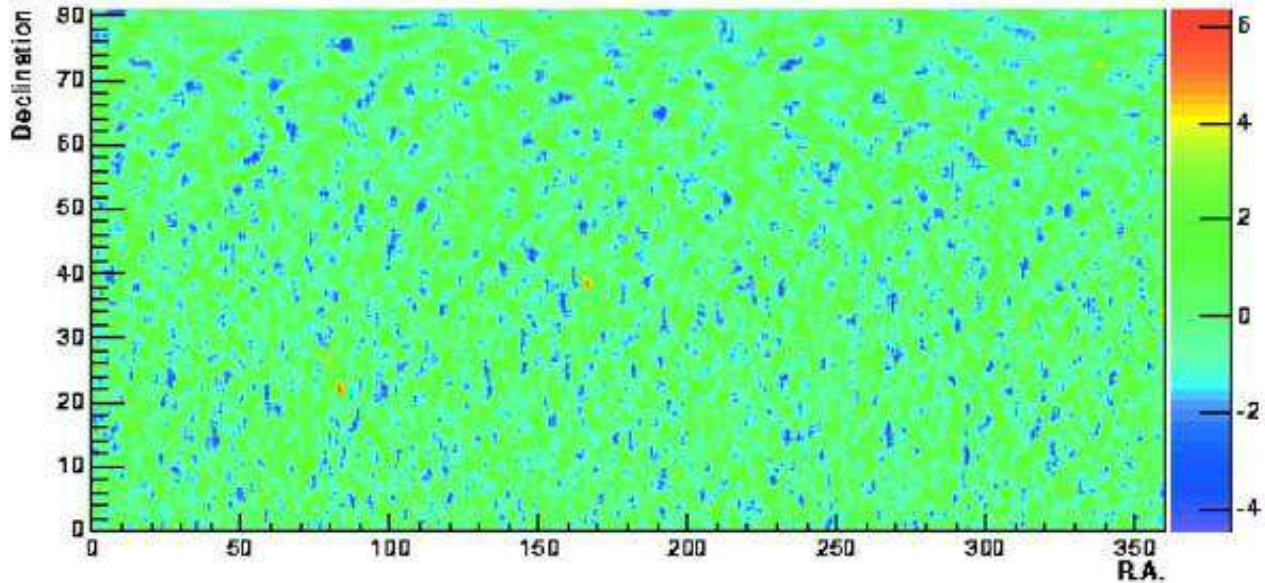}
\caption{Northern hemisphere as seen in TeV gamma rays. At each point, the excess is 
summed over a 2.1 by [$2.1/\cos(\delta)$]$^{\circ}$ bin, and the significance of the 
excess in standard deviations is shown by the color scale (Figure from~\cite{atkins04}).} \label{all-sky-map}
\end{figure*}

A full survey of the northern hemisphere (declination of 1.1--80$^{\circ}$) for point
sources was carried out~\citep{atkins04}. Figure~\ref{all-sky-map} shows the map of the northern 
hemisphere in TeV gamma rays for the Milagro data set between 15 December 2000 and 25 
November 2003. Table~\ref{table1} lists all point sources with an excess greater 
than 4$\sigma$. The last column represents the 95\% confidence upper limit on the flux, in 
units of the Crab. The Crab and Mkn 421 are both clearly visible in the map, and are listed in 
the sixth and seventh row, respectively, of Table~\ref{table1}. A square bin of $2.1^{\circ}$ in 
declination ($\delta$) and [2.1/cos($\delta$)]$^{\circ}$ in right ascension was used in this 
analysis. To detect other possible sources of VHE emission it is necessary to survey the sky at 
many different timescales and at many different bin sizes. In addition to searching for steady 
point sources of TeV gamma rays, Milagro has searched for short bursts of TeV gamma 
rays~\citep{2004ApJ...604L..25A}, for TeV emission from the direction of satellite detected 
GRBs~\citep{saz04}, and for extended sources of TeV emission~\citep{smith04}, as we summarize in 
this paper.

\begin{table}[t]
\begin{center}
\caption{Locations of All Regions with an Excess Greater than 4$\sigma$}
\begin{tabular}{|c|c|c|c|c|c|c|}
\hline \textbf{R.A.} & \textbf{Decl.} & \textbf{ON} & \textbf{OFF} & \textbf{Excess} & \textbf{$\sigma$} & \textbf{UL}
\\
\hline

0.3 &	34.3&	3.12308e+06&	3.11456e+06&	8623&	4.7&	0.84	\\
37.8 &	6.7&	7.02166e+05&	6.98667e+05&	3498&	4.0&	1.8	\\
43.6 &	4.8&	5.85952e+05&	5.82716e+05&	3236&	4.1&	2.0	\\
49.1 &	22.5&	2.21431e+06&	2.20813e+06&	6175&	4.0&	0.87	\\
79.9 &	26.8&	2.57841e+06&	2.57025e+06&	8161&	4.9&	0.97	\\
83.6 &	22.0&	2.17188e+06&	2.16222e+06&	9665&	6.3&	NA	\\
166.5 &	38.6&	3.23552e+06&	3.22467e+06&	10850&	5.8&	NA	\\
306.6 &	38.9&	3.25329e+06&	3.24531e+06&	7983&	4.2&	0.78	\\
313.0 &	32.2&	3.08380e+06&	3.07548e+06&	8320&	4.5&	0.85	\\
339.1 &	72.5&	6.63534e+05&	6.59727e+05&	3807&	4.2&	3.02	\\	
356.4 &	29.5&	2.98656e+06&	2.97910e+06&	7455&	4.1&	0.84	\\

\hline
\end{tabular}
\label{table1}
\end{center}
\end{table}

\section{THE GALACTIC PLANE}
\begin{figure}[t]
  \begin{center}
    \begin{tabular}{c}
\includegraphics[width=80mm]{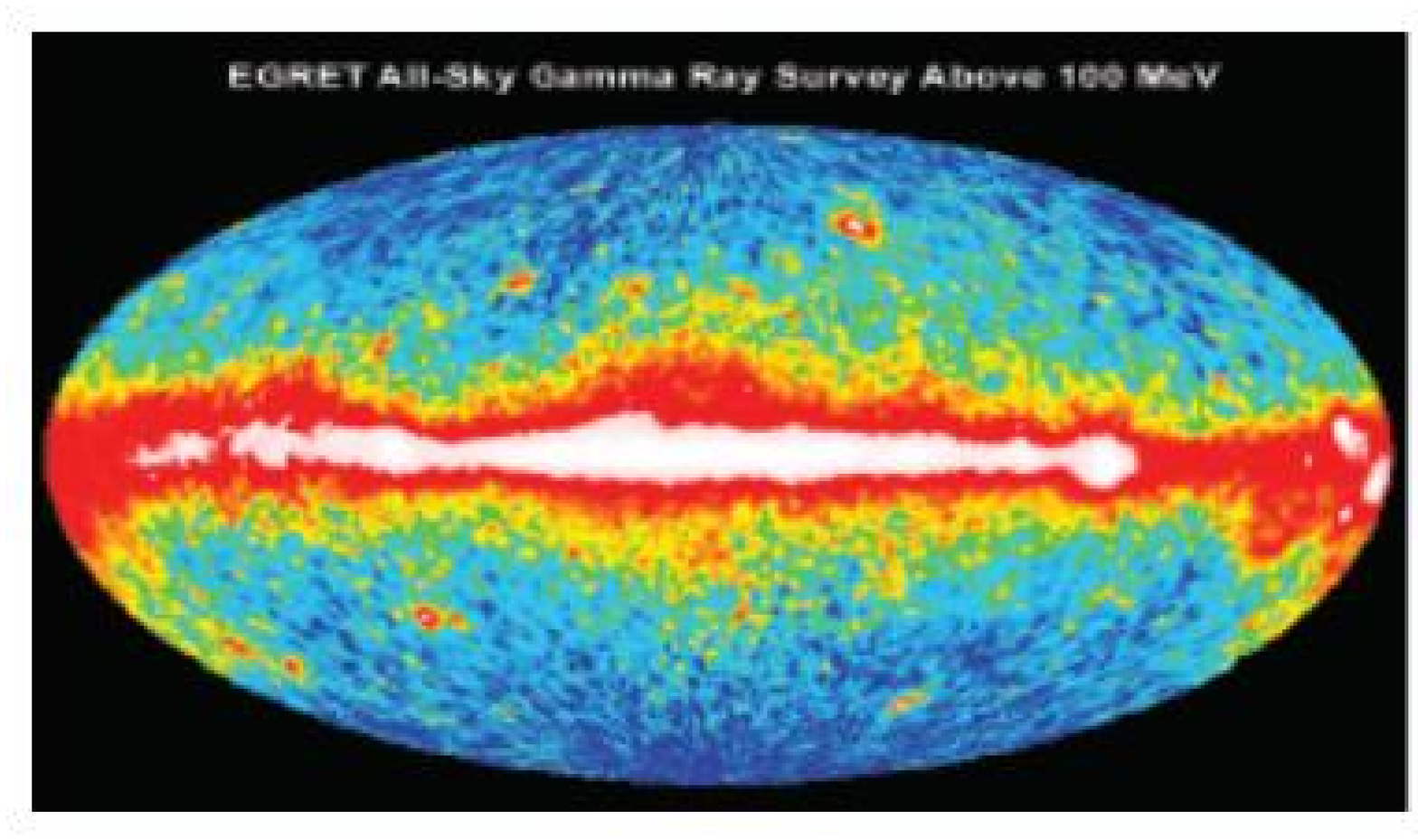} \\
\includegraphics[width=80mm]{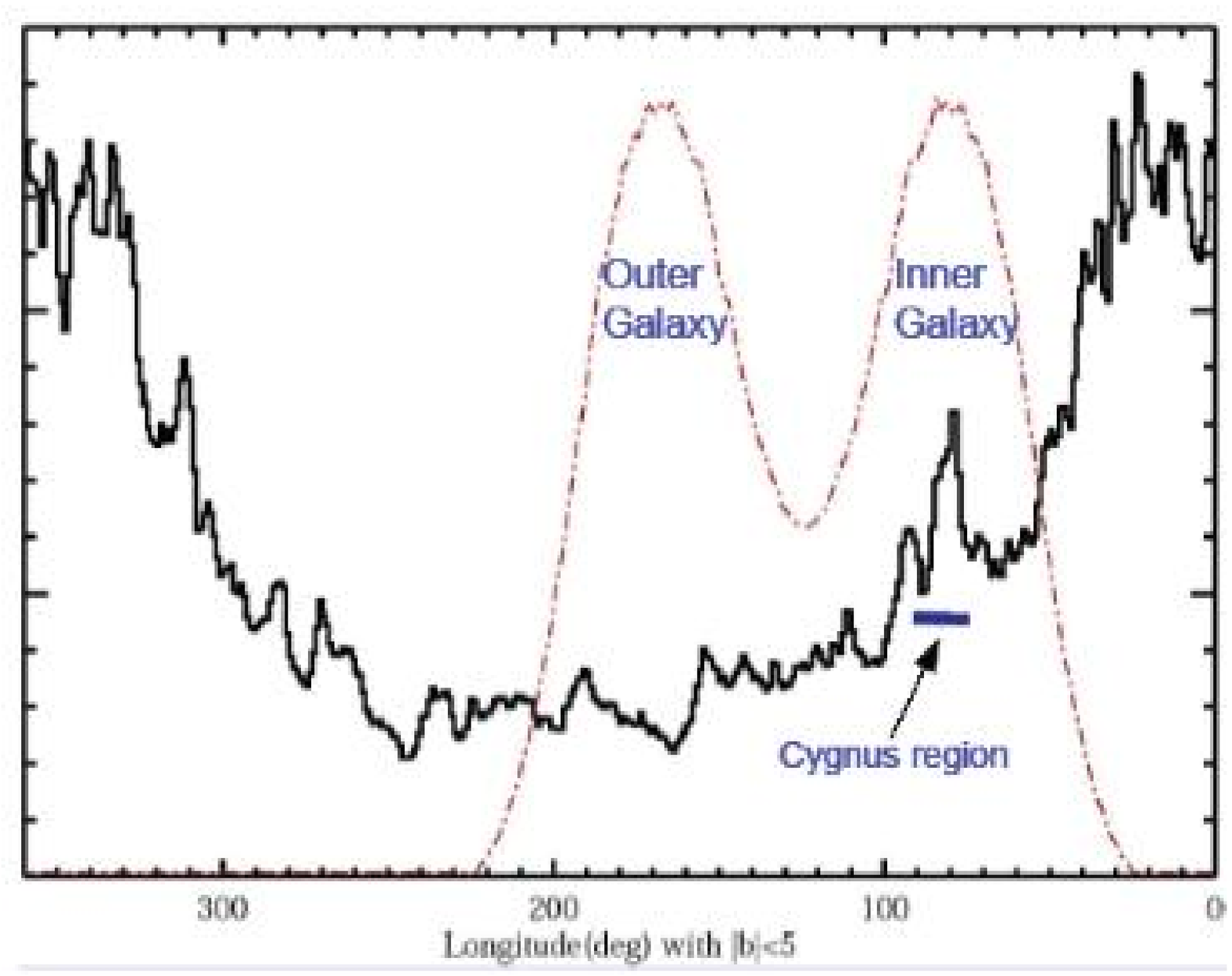} \\
    \end{tabular}
\caption{{\bf Top} -- EGRET all-sky map showing Galactic plane emission above 100 MeV. 
{\bf Bottom} -- EGRET diffuse GeV flux (in black) along with the Milagro exposure (in red).}
\label{egret_figure}
\end{center}
\end{figure}

\begin{figure*}[t]
  \begin{center}
    \begin{tabular}{cc}
\includegraphics[width=85mm]{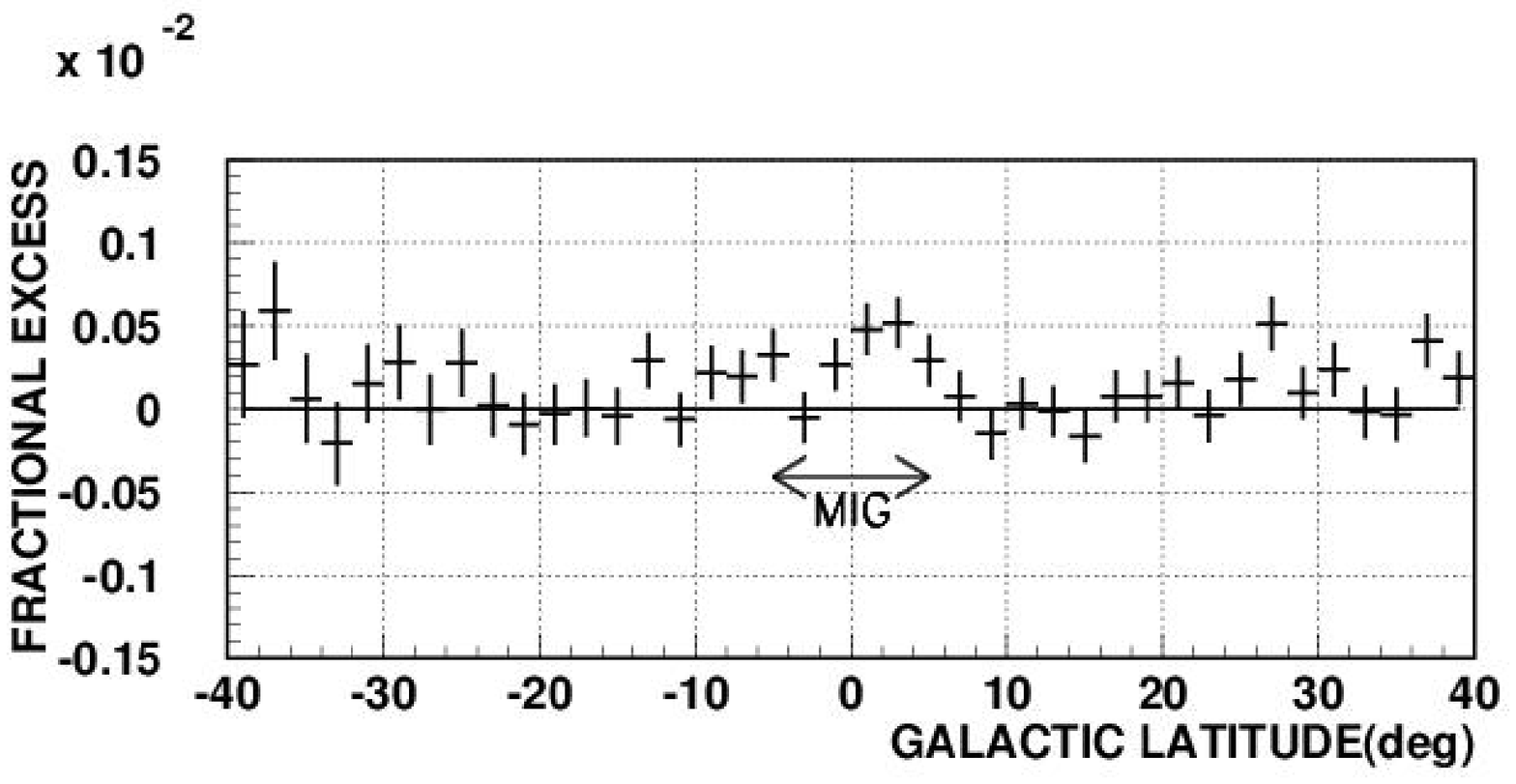} &
\includegraphics[width=85mm]{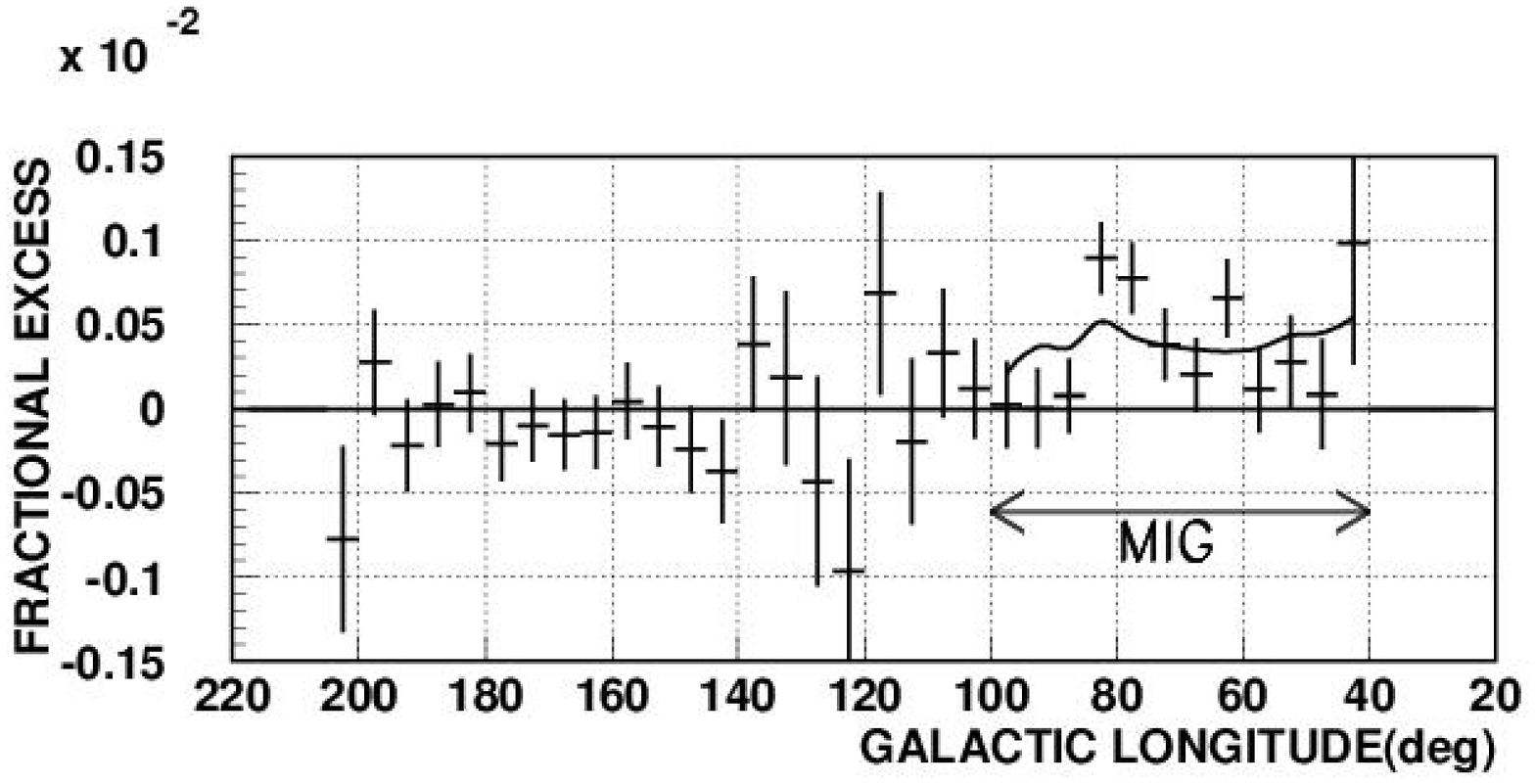} \\
    \end{tabular}
\caption{{\bf Left} -- Profile of the fractional excess as a function of Galactic latitude (
for a Galactic longitude between 40 and 100 degrees). {\bf Right} -- Profile of the
fractional excess as a function of Galactic longitude (for a Galactic latitude between
-5 and 5 degrees). The EGRET longitudinal source shape is superposed. (Figures from ~\cite{atkins05})}
\label{profile}
\end{center}
\end{figure*}

\begin{figure*}[t]
\centering
\includegraphics[width=150mm]{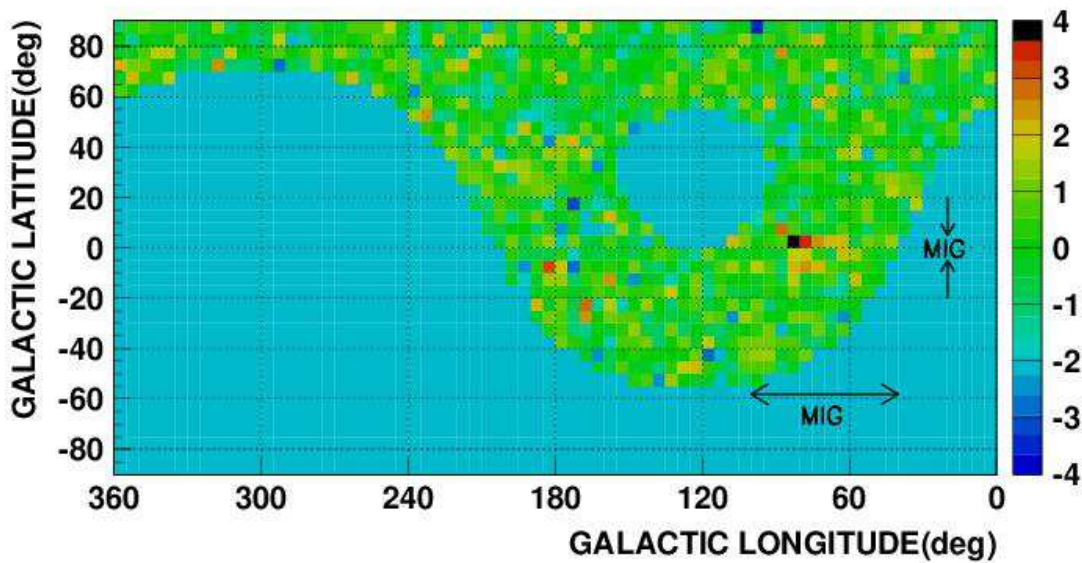}
\caption{Map in Galactic coordinates of Milagro significances 
in $5^{\circ}$ by $5^{\circ}$ bins. The light area has no Milagro exposure (Figure from ~\cite{atkins05}).} \label{galaxy}
\end{figure*}

Diffuse emission from the Galactic plane is the dominant source in the gamma-ray 
sky~\citep{hunter97}. Most of the diffuse VHE emission from the Galactic plane 
is thought to be produced by the interaction of cosmic-ray hadrons with the interstellar 
matter. The flux measured by EGRET below 1 GeV fits models well, but that measured 
between 1 and 40 GeV is significantly larger than what is predicted by most models. One 
possible explanation for this enhanced emission is the inverse-Compton scattering of 
cosmic-ray electrons~\citep{1997JPhG...23.1765P}. If this turns out to be the dominant 
source of diffuse gamma-ray emission from the Galactic plane, then the flux at TeV 
energies could be an order of magnitude higher than previously thought. Figure~\ref{egret_figure} 
shows the EGRET observations of the Galactic plane. The top panel shows the all-sky
map, in Galactic coordinates, produced by EGRET, clearly showing the Galactic plane. The 
bottom panel shows the EGRET diffuse GeV flux (in black) along with the Milagro exposure 
(in red), indicating that the Milagro observation region is optimized for the detection of the diffuse 
emission predicted by the EGRET measurements. Using 36 months of data, from 19 July 2000 
to 18 July 2003, we looked at the inner (40-100 degrees) and outer (140-220 degrees) 
regions of the Galaxy~\citep{fleysher04}. While the outer Galaxy shows no significant 
excess, the inner Galaxy shows a 5 sigma excess~\citep{fleysher04}. Figure~\ref{profile} shows the 
profile in latitude for the longitude band (40 to 100 degrees) of the inner Galactic 
region (left panel) and the profile in longitude for the latitude band (-5 to 5 degrees) of 
the inner Galactic region (right panel), where the enhancement can be seen just north of the 
equator. Figure~\ref{galaxy} shows the significance map of the Galaxy. The region of the inner 
Galaxy shows an enhancement along and just north of the Galactic equator. This is the same region 
where EGRET detected the strongest signal in the 100 MeV energy range. The 5 sigma excess 
is seen by summing the entire inner Galaxy with a +/- 5 degree latitude band, as was 
suggested by the EGRET results. We note that the Milagro observation of the Galactic 
plane remains significant even when the region around the Cygnus Arm is excluded. 
This constitutes the first detection of the Galactic plane at TeV energies.

\section{EXTENDED SOURCES}

A search for extended emission was carried out for the Milagro data collected between 
17 August 2000 and 5 May 2004~\citep{smith04}. A set of standard cuts has been developed by 
the Milagro collaboration and validated by observations of the Crab~\citep{atkins03} 
and Mkn 421~\citep{atkins04}. Events with 20 or more PMT hits used by the angle fitter (NFIT $\geq$ 20)
are kept (this rejects about 20\% of events which have poor fits). In addition, we cut on a parameter 
known as ``compactness''~\citep{atkins04} (X2$\geq$2.5) to retain 50\% of the gammas while 
removing more than 90\% of the background protons. The excess at each position is determined by counting 
the number of events in a particular bin and subtracting the estimated background. 
The background is computed from data collected at the same local detector coordinates, 
but at a different time, ensuring the celestial angles of the background event sample 
do not overlap with the source position under consideration. The method of \cite{lima} is 
used to compute the significance of each excess.
While the optimal square bin for detection of point sources with Milagro is 2.1 degrees 
on each side~\citep{atkins04}, to look for diffuse sources, the standard Milagro 
sky maps were searched using a range of bin sizes from 2.1 to 5.9 degrees in steps of 
0.2. 20 separate searches were performed on the same maps, though the results are highly 
correlated. Monte Carlo simulations were used to compute the post-trials probability for 
each source candidate.

\subsection{3EG J0520+2556}
\begin{figure*}[t]
  \begin{center}
    \begin{tabular}{cc}
\includegraphics[width=85mm]{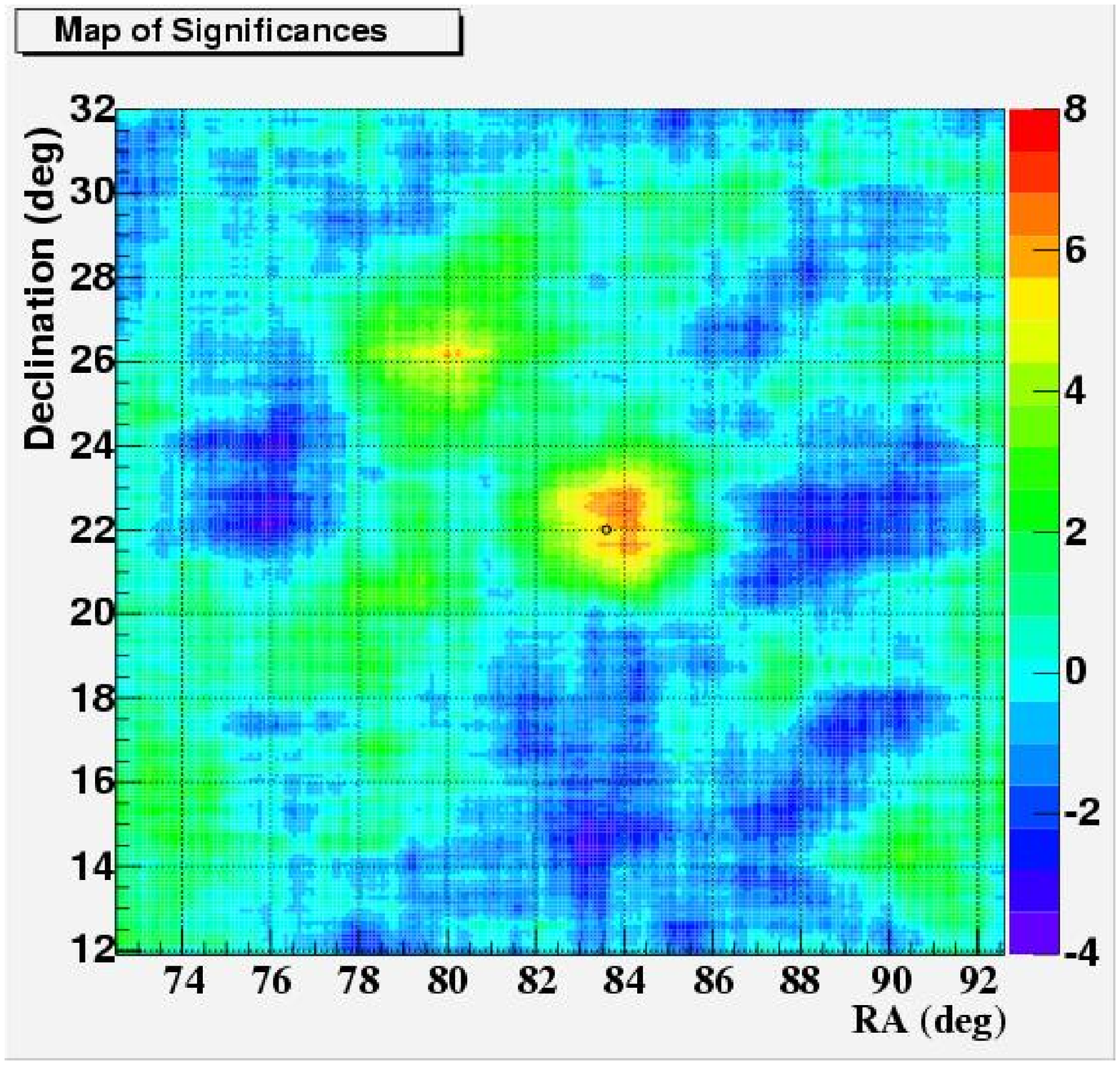} &
\includegraphics[width=85mm,height=82mm]{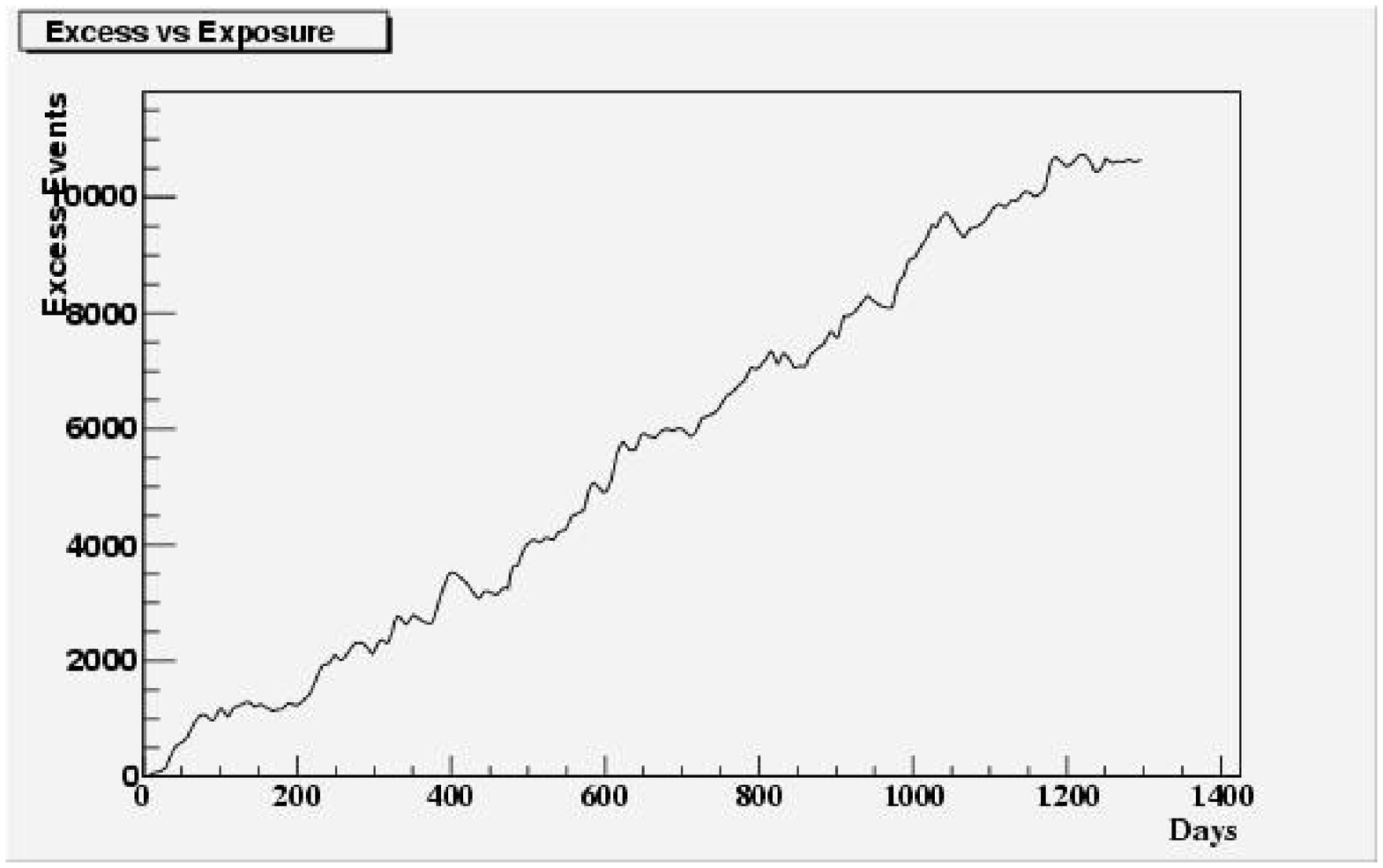} \\
    \end{tabular}
\caption{{\bf Left} -- Milagro significance map, showing the Crab in the center and the TeV source
coincident with EGRET source 3EGJ0520+2556 to the left. {\bf Right} -- Accumulated significance
on TeV 0520+2556 as a function of time (Figure from~\cite{smith04}).}
\label{egret_both}
\end{center}
\end{figure*}

\begin{figure}[t]
\centering
\includegraphics[width=70mm]{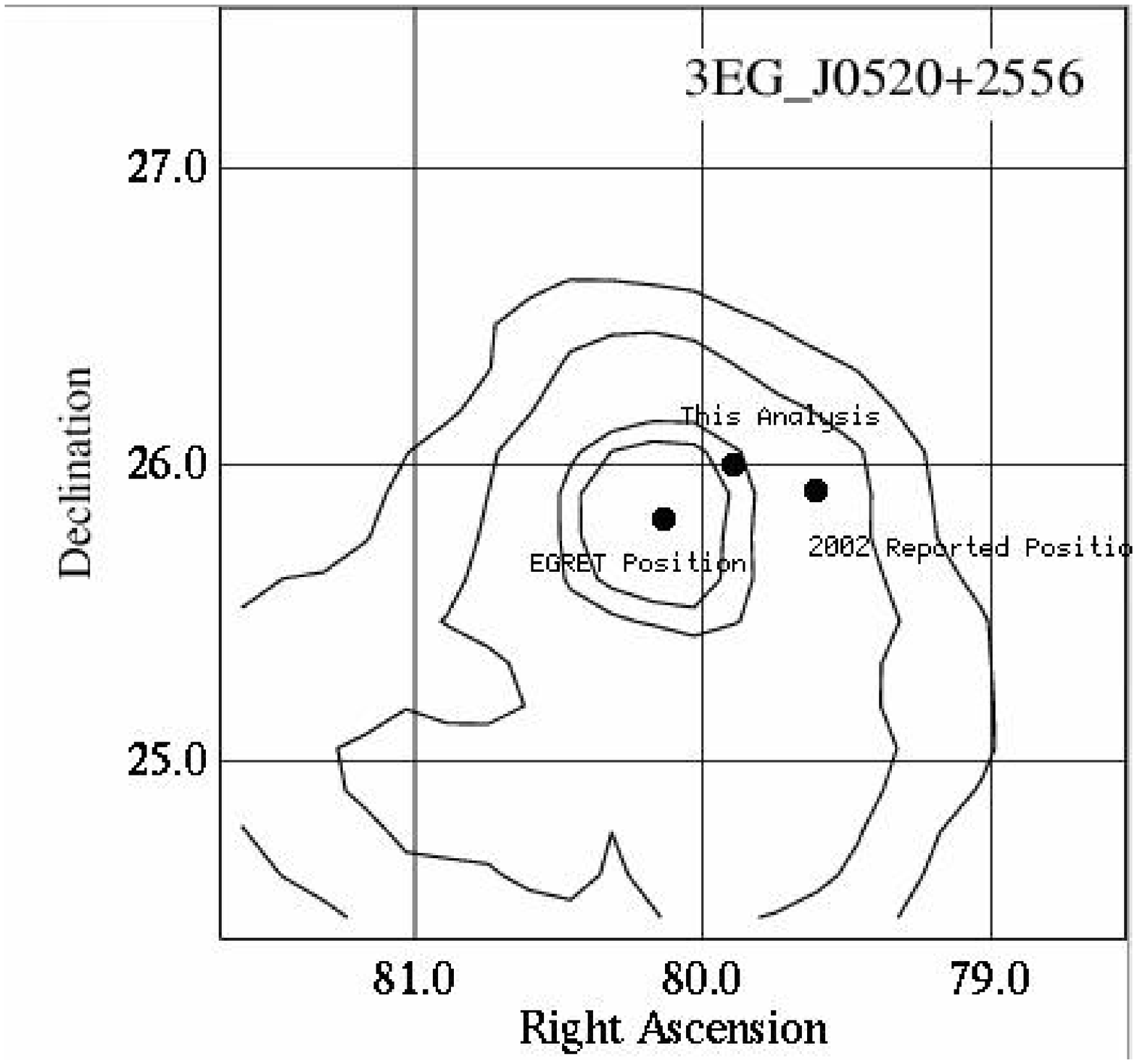}
\caption{EGRET Source Location and contours. We include the position reported
by Milagro in 2002~\citep{sinnis02}, as well as the current location of maximum significance (Figure from~\cite{smith04}).} \label{egret_source}
\end{figure}

The most significant candidate found in our search of extended sources had a pre-trials 
significance of 5.9 sigma, located at RA=79.8 degrees and Dec=26.0 degrees and was identified 
using a 2.9 degree bin size. The probability of observing an excess this significant at any point 
in the sky at any bin size is 0.8\%. Figure~\ref{egret_both} (left panel) shows the map of 
significances around the source, which is located $\sim$5.5 degrees from the Crab. This 
candidate was first reported in 2002~\citep{sinnis02}. The cumulative significance using only data 
since it was first reported is 3.7$\sigma$. The right panel of 
Figure~\ref{egret_both} shows the accumulation of significance with time, indicating that 
the excesss increases at a constant rate and shows no periods of significant flaring. This 
candidate is coincident with the EGRET unidentified source 3EG J0520+2556 
(See Figure~\ref{egret_source}).

\subsection{The Cygnus Arm}

\begin{figure*}[t]
\centering
\includegraphics[width=130mm]{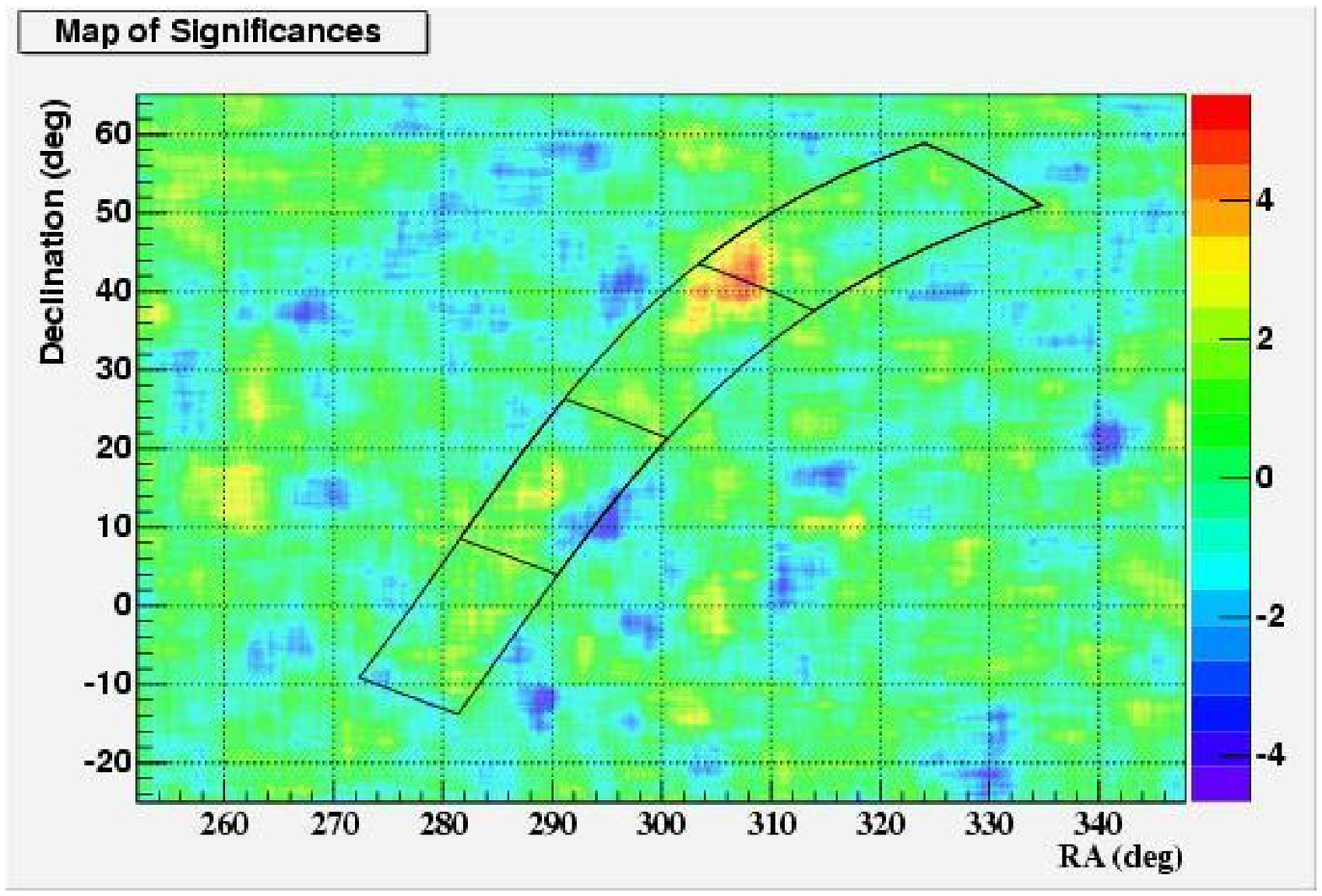}
\caption{Milagro significance map, showing a clear excess in the Cygnus region. Also
shown is the Galactic plane region from a latitude of 20 degrees to 100 degrees and
longitude of -5 to 5 degrees (Figure from~\cite{smith04}).} \label{cygnus_region}
\end{figure*}

The second extended source candidate is coincident with the region known as the Cygnus Arm, 
a spiral arm within our Galaxy that extends radially away from observers on Earth. This 
region is known to be a dense region of gas and dust and was observed by EGRET as the 
brightest source of GeV gamma rays in the northern sky, with a 
diffuse GeV emission comparable to the Galactic bulge. Like the emission from the Galactic 
plane region, VHE emission from the Cygnus Arm is thought to originate mainly from 
interactions of cosmic rays with the interstellar gas and dust. A 5.5 sigma excess was 
detected using a 5.9 degree bin, at RA=308 degrees and Dec=42 degrees. The probability of 
observing an excess this significant at any point in the sky at any bin size is 2.0\%. 
Figure~\ref{cygnus_region} shows the significance of the region of the sky containing the 
Cygnus arm, showing also the Galactic plane from l=20 degrees to l=100 degrees and 
b=+/- 5 degrees superimposed on the plot. The excess observed by Milagro is inconsistent 
with a point source and the number of events coming from the entire 5.9$^{\circ}$ bin is 
approximately twice that of the Crab. Like in the case of 3EG J0520+2556, the 
accumulation of the excess is steady, and no evidence for flaring is observed. While this 
is an extremely bright region, making it the hottest spot in the Galactic plane, it is not 
surprising that it has not been detected yet by any of the Atmospheric Cerenkov Telescopes, given 
the diffuse nature of the source, and the limited field of view of such 
telescopes.

\section{CONCLUSION}

Milagro has detected three previously unknown diffuse sources of TeV gamma rays.
Milagro has reported the first observation of diffuse TeV emission from the Galactic plane.
In addition to this, two other diffuse sources have been detected. The first, 
coincident with 3EG J0520+2556, and the second coincident with the Cygnus Arm, at 5.9 and 
5.5 sigma respectively. When all trials of the all-sky search are considered, the 
probabilities of these excesses being due to background fluctuations are 0.8\% and 2.0\% 
respectively. The source coincident with 3EG J0520+2556 was previously reported by 
Milagro as a ``hot spot''.

%





\bigskip 
\begin{acknowledgments}
Many people helped bring Milagro to fruition.  In particular, we
acknowledge the efforts of Scott DeLay, Neil Thompson and Michael Schneider. 
This work has been supported by the National Science Foundation (under grants 
PHY-0075326, 
-0096256, 
-0097315, 
-0206656, 
-0245143, 
-0245234, 
-0302000, 
and
ATM-0002744) 
the US Department of Energy (Office of High-Energy Physics and 
Office of Nuclear Physics), Los Alamos National Laboratory, the University of
California, and the Institute of Geophysics and Planetary Physics.
\end{acknowledgments}

\bigskip 

\end{document}